\documentclass[twocolumn,floatfix,prl,showpacs,floatfix,superscriptaddress]{revtex4}
\usepackage{graphicx}
\usepackage{color}
\usepackage{psfrag}
\usepackage{epsfig}
\usepackage{bbm}
\usepackage{bm}
\newcommand{\ket}[1]{| #1 \rangle}
\newcommand{\bra}[1]{\langle #1 |}
\newcommand{\rb}[1]{\left( #1 \right)}

\newcommand{\ew}[1]{\langle #1 \rangle}
\newcommand{\beq}{\begin{eqnarray}}
\newcommand{\eeq}{\end{eqnarray}}

\begin{document}
\title{Fast initialization of the spin state of an electron in a
quantum dot in the Voigt configuration}
\author{C. Emary}
\affiliation{Department of Physics,
        The University of California-San Diego,
        La Jolla, CA 92093}
\author{Xiaodong Xu}
\author{D. G. Steel}
\affiliation{The H. M. Randall Laboratory of Physics,
            The University of Michigan,
            Ann Arbor, MI 48109}
\author{S. Saikin}
\author{L. J. Sham}
\affiliation{Department of Physics,
        The University of California-San Diego,
        La Jolla, CA 92093}

\date{August 9th 2006}
\begin{abstract}
  We consider the initialization of the spin-state of a single electron
  trapped in a self-assembled quantum dot via optical pumping of a trion level.
  We show that with a magnetic field applied perpendicular to the
  growth direction of the dot, a near-unity fidelity can be obtained
  in a time equal to a few times the inverse of the spin-conserving
  trion relaxation rate.  This method is several orders-of-magnitude
  faster than with the
  field aligned parallel, since this configuration must rely on a slow hole
  spin-flip mechanism.  This increase in speed does result in a
  limit on the maximum obtainable fidelity, but we show that for
  InAs dots, the error is very small.
\end{abstract}
\maketitle

In a recent experiment \cite{ata06}, Atat\"ure and coworkers
demonstrated high fidelity spin-state preparation of an electron in
a self-assembled InAs/GaAs quantum dot (QD).
Their purifying mechanism coupled the resonant laser excitation of
the QD with heavy-light hole mixing to induce a small, but finite,
degree of spin-flip Raman scattering.
These experiments were performed with the magnetic field aligned in
the growth direction of the QD (the Faraday configuration), and this
was shown to be effective in suppressing deleterious spin-flips
caused by the nuclear hyperfine field.

Although fidelities very close to unity ($\ge 99.8$\%) were obtained
through this mechanism, for quantum information processing purposes
one would also like state preparation to be fast. The speed of the
scheme in Ref.~\cite{ata06} is limited by the rate of hole-mixing
spin-flip trion relaxation, which was determined to have a
characteristic time of $\sim 1~\mu$s, corresponding to the measured
rate of $100$~kHz.
This is slow compared with the picosecond timescale on which it is
hoped that quantum operations will be performed in such dots
\cite{pie02,cal03,lov05,che04,ce06}.

It is the purpose of this paper to show that a magnetic field
aligned perpendicular, rather than parallel, to the growth axis
allows the purification of the spin to near-unity fidelities with a
characteristic time-scale of $2 \Gamma^{-1} \approx 1$~ns, where
$\Gamma=300$~MHz is the trion relaxation rate without spin-flip as
measured by Atat\"ure. This Voigt configuration is therefore some
three orders-of-magnitude quicker than the Faraday configuration of
Ref.~\cite{ata06}.

The price paid for this dramatic speed-up is that now both ground
states are optically coupled to the trion.
This inevitably leads to a reduction in the maximum obtainable
fidelity, as it provides a path back for the population localized in
the desired level.
However, as we will show, this effect decreases with increasing
field strength such that, for a typical InAs QD, the maximum
obtainable fidelity typically differs from unity by only $0.3$\% at
a field of $1$~T and $0.005$\% at $8$~T.

\begin{figure}[t]
  \begin{center}
  \psfrag{t+}{{\Large$\ket{\tau+}$}}
  \psfrag{t-}{{\Large$\ket{\tau-}$}}
  \psfrag{x+}{{\Large$\ket{x+}$}}
  \psfrag{x-}{{\Large$\ket{x-}$}}
  \psfrag{Eze}{{\Large$E_B^e$}}
  \psfrag{Ezh}{{\Large$E_B^h$}}
  \psfrag{V1}{{\Large$V_1$}}
  \psfrag{V2}{{\Large$V_2$}}
  \psfrag{H1}{{\Large$H_1$}}
  \psfrag{H2}{{\Large$H_2$}}
  \epsfig{file=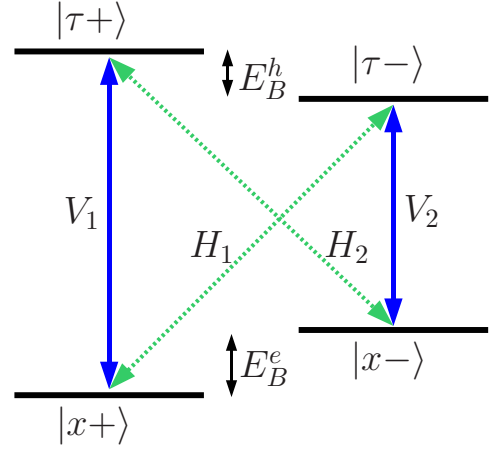, clip=true,width=0.75\linewidth}
  \caption{
    The four levels of the electron-trion system in the Voigt basis consists of
    two Zeeman-split single-electron ground states $\ket{x \pm}$
    with spins in the $x$ direction,
    and two trion levels $\ket{\tau \pm}$ with
    heavy-hole spins also in the $x$ direction.
    Arrows indicate allowed optical transitions with $H$, $V$ denoting
    two orthogonal linear polarizations.
    State-preparation is achieved by resonantly pumping the $V_1$
    transition.  This populates the trion level $\ket{\tau_+}$,
    which subsequently relaxes
    with rate $\Gamma$ back to both ground states, resulting in a
    partial
    transfer of population from $\ket{x+}$  to $\ket{x-}$.  This
    simple picture is complicated by the fact that the same laser also
    drives the $V_2$ transition, albeit off-resonantly.  This
    results in a small pumping of population in the opposite direction and hence
    a slight decrease in the maximum obtainable purity.
    We take the relaxation rate $\Gamma=1.2~\mu$eV and
    $g$-factors $g^e_x=-0.46$ and  $g^h_x=-0.29$.
    For a field of 1T, the Zeeman splitting are then $E^e_B = -27~\mu$eV
    and $E^h_B = 17~\mu$eV.
    \label{f1}
 }
  \end{center}
\end{figure}

We consider a singly-charged self-assembled InAs QD with growth
direction $z$.  Figure \ref{f1} shows our four-level model that
describes the pertinent features of the system.
With $B$-field aligned in the $x$-direction, the Zeeman energy of a
QD electron is ${\cal H}_B^e = g_x^e \mu_B B_x s^e_x \equiv E^e_B
s^e_x$, where $g_x^e$ is the electronic $g$-factor, $\mu_B$ is the
Bohr magneton,  $B_x$ is the magnitude of the field, and  $s^e_x =
\pm 1/2$ corresponds to the electron spin.  We have measured the
magnitude of the electron $g$-factor to be $|g_x^e| = 0.46$
\cite{xia06}, which is similar to values in the literature
\cite{ata06,hap02}.

Our measurements also indicate that the behaviour of the heavy-hole
component of the trion in this field can be described with a Zeeman
Hamiltonian ${\cal H}_B^h = - g_x^h \mu_B B_x s^h_x \equiv E^h_B
s^h_x$, where $s^h_x=\pm 1/2$ are the eigenvalues of a pseudo-spin,
the components of which correspond to heavy-hole states aligned in
the $x$-direction, and $g_x^h$ is the hole $g$-factor, which we
determine to have a magnitude of $|g_x^h|=0.29$.  Our measurements
do not give us access to the signs of these two $g$-factors, but
here we take both to be negative as suggested by some recent results
\cite{ata06,bay00}. Our scheme relies neither on this assumption
about the signs, nor indeed even on $g_x^h$ being non-zero.

The four levels of our model are then: the two electron ground
states with spins in the $x$-direction, $\ket{x\pm} \equiv
2^{-1/2}\rb{\ket{\downarrow} \pm \ket{\uparrow}}$, where
$\ket{\downarrow}$ and $\ket{\uparrow}$ represent electron spins in
the $z$ direction; and the two trion levels, $\ket{\tau\pm} \equiv
2^{-1}\rb{\ket{\downarrow\uparrow}-
\ket{\uparrow\downarrow}}\rb{\ket{\Downarrow} \pm \ket{\Uparrow}}$,
where $\ket{\Downarrow}$ and $\ket{\Uparrow}$ denote heavy-hole
states also aligned in the $z$ direction.
Figure \ref{f1} shows the allowed optical transitions between these
levels.  These transitions are linearly polarized and we have
defined the polarization vectors in terms of $\sigma_\pm$
circular polarizations as $V = 2^{-1/2}(\sigma_- + \sigma_+)$ and
$H = 2^{-1/2}(\sigma_- - \sigma_+)$.
We drive the system with a $V$-polarized laser tuned on resonance
with the transition from $\ket{x+}$ to $\ket{\tau+}$, which is
denoted $V_1$ in Fig.~\ref{f1}. This illumination will also drive
the $V_2$ transition and this off-resonant driving is the main
source of non-ideality considered in our model.  We elect to drive
the $V_1$ transition because, since we take the sign of both
electron and hole $g$-factors to be the same, the detuning of $V_2$
with respect to driving transition $V_1$ is $\Sigma_B = (g_x^e +
g_x^h) \mu_B B_x = E_x^e - E_x^h $.  The magnitude of this detuning
is greater than that of $\Delta_B = (g_x^e - g_x^h) \mu_B B_x =
E_x^e + E_x^h $, which is the detuning of transition $H_2$ with
respect to transition $H_1$.  As we will show, the larger this
detuning, the smaller the deleterious effects of the off-resonant
transition \cite{note_gs}.

In the rotating frame then, the Hamiltonian of our system with
driven $V_1$ transition in the basis $\left\{ \ket{x+},
\ket{x-},\ket{\tau+},\ket{\tau_-} \right\}$ is
\beq
  {\cal H} =
  \rb{
    \begin{array}{cccc}
      0 & 0 & \Omega & 0\\
      0 & 0 & 0 & \Omega e^{- i\Sigma_B t} \\
      \Omega & 0 & 0  & 0 \\
      0 & \Omega e^{i \Sigma_B t} & 0 & 0 \\
     \end{array}
     \label{Ham1}
  },
\eeq
where $\Omega$ is the Rabi frequency of the laser and we have set
$\hbar=1$. As this Hamiltonian shows, the laser drives not only the
transition with which it is resonant, but also the unintended
transition with terms oscillating with frequency $\pm \Sigma_B$.
In writing this Hamiltonian, we have neglected hole-mixing since it
is both expected to be small~\cite{ata06}, and can in any case be
incorporated into the current scheme without significant change
\cite{mix_note}.


We will determine the properties of this system through the master
equation for the density matrix $\rho$ in the Lindblad form
\beq
  \dot{\rho} = -i \left[{\cal H},\rho\right] +\sum_i {\cal L}_i[\rho]
  \label{master}
  ,
\eeq
where the sum is over all trion relaxation channels, each of which
is described by a Lindblad superoperator
\beq
  {\cal L}_i[\rho] = D_i \rho D_i^\dag
  - \frac{1}{2} D_i^\dag D_i \rho
  - \frac{1}{2} \rho D_i^\dag D_i
  .
\eeq
Since the trion can relax through all four optical transitions shown
in Fig.~\ref{f1}, we need to consider the four independent jump
operators: $
  D_1 = \sqrt{\Gamma} \ket{x+}\bra{\tau+}
$, $
  D_2 = \sqrt{\Gamma} \ket{x-}\bra{\tau+}
$, $
  D_3 = \sqrt{\Gamma} \ket{x+}\bra{\tau-}
$, and $
  D_4 = \sqrt{\Gamma} \ket{x-}\bra{\tau-}
$.
In writing these operators, we have assumed that the relaxation
channels proceed incoherently. This is justified since we will work
in a regime where the Zeeman splittings are large enough that
$|\Delta_B|, |\Sigma_B| > \Omega$ and the degree of spontaneously
generated coherence \cite{eco05} is negligible. We also assume for
simplicity that the rate $\Gamma$ is the same for all channels. On
the time-scales considered here, the hole-mixing spin-flip
relaxation, central to the mechanism of Ref~\cite{ata06}, is
negligible. Furthermore, since we will work at significant magnetic
fields ($\gtrsim 1$~T), the nuclear hyperfine interaction is frozen
out and can also be neglected.
Finally, we assume that the initial state of the spin is unpolarized
with $\rho_{++}=\rho_{--}=1/2$ and all other elements of $\rho$
zero.


Our elucidation of the properties of this system consists of two
parts. Firstly, we derive the time taken for the system to reach its
asymptotic limit.  This we do by neglecting the effects of the
off-resonant transition. Secondly, we include the off-resonant
effects and derive a limit on the maximum fidelity obtainable
imposed by this nonideality.

With a trion relaxation rate of $\Gamma=1.2~\mu$eV and $g$-factors
as stated, then even with a small applied magnetic field, we work in
a regime in which the detuning of the off-resonant transition
$|\Sigma_B|$ is much greater than both $\Gamma$ and the Rabi energy
$\Omega$. In this limit we can assume that the terms $\Omega e^{\pm
i \Sigma_B t}$ in the Hamiltonian of Eq.~(\ref{Ham1}) oscillate
sufficiently rapidly that they approximate as self-averaging to
zero. In this case, the state $\ket{\tau-}$ decouples from the rest
of the system and the Hamiltonian reduces to ${\cal
H}=\Omega\ket{x+}\bra{\tau+}+\mathrm{H.c.}$. Physically this means
that the off-resonant transition is so far off resonance that the
laser induces no transitions from it.  We will derive a correction
to this behaviour later.
\begin{figure}[t]
  \begin{center}
  \psfrag{TG}{$T_0 \Gamma$}
   \psfrag{r}{$\Omega/\Gamma$}
  \epsfig{file=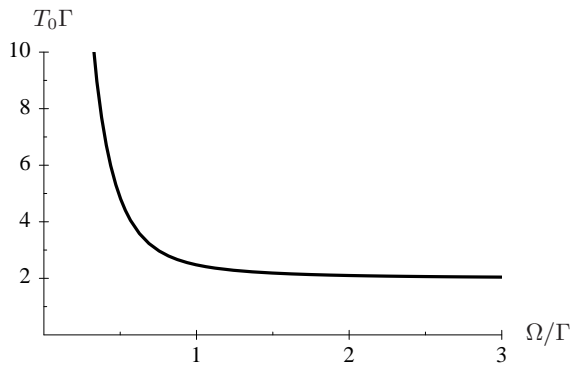, clip=true,width=0.9\linewidth}
  \caption{The characteristic time $T_0$ describing the approach of
    the system to its asymptotic limit
    as a function of the Rabi frequency $\Omega$.
    When $\Omega/ \Gamma \gtrsim 1$, $T_0\approx 2/\Gamma$.
    Parameters as in Fig~\ref{f1}.
    \label{T0fig}
 }
  \end{center}
\end{figure}

With this simplified Hamiltonian, there are only three independent
non-zero density matrix elements to consider and these we organize
into the vector  $ \mathbf{v} =
\rb{\rho_{x+,x+},\rho_{x-,x-},\mathrm{Im}\rho_{x+,\tau+}}$.  We have
utilized the normalization condition $1=\mathrm{Tr}\rho$ to
eliminate $\rho_{\tau+,\tau+}$.

The equation of motions for these components can then be rephrased
in terms of this vector as
\beq
  \dot{\mathbf{v}} = \mathbf{X}\cdot \rb{\mathbf{v}-\mathbf{v}_\infty}
  \label{vEoM}
  ,
\eeq
where
\beq
  X=
  \rb{
    \begin{array}{ccc}
      -\Gamma & -\Gamma & -2\Omega \\
      -\Gamma & -\Gamma & 0   \\
     2\Omega & \Omega   &  -\Gamma
     \end{array}
  }
  \label{X}
  ,
\eeq
and $\mathbf{v}_\infty = \rb{0,1,0}$ is the stationary solution of
this model, which represents the qubit population completely
localized in state $\ket{x-}$ and hence 100\% purified.

\begin{figure}[t]
  \begin{center}
  \psfrag{t}{$t$~(ns)}
  \psfrag{rho}{$\rho_{--}$}
  \epsfig{file=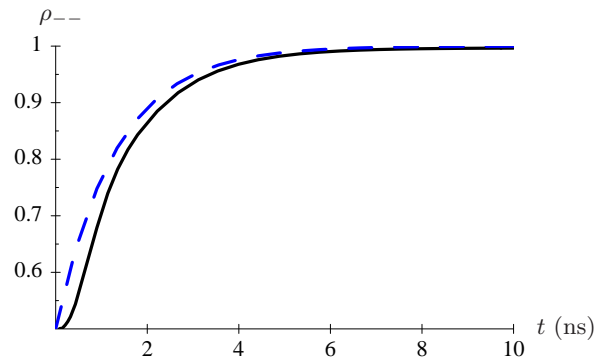, clip=true,width=0.9\linewidth}
  \caption{ The population of the state $\ket{x-}$ as a function of
    time under continuous illumination.  The solid black line shows
    the result of numerical integration of the master equation,
    whereas the dashed blue line shows the analytic result of
    Eq.~(\ref{vinfty}).
    Near-unity fidelity is approached with a characteristic time of
    $T_0\approx 1$~ns.
    The parameters here are the same as Fig.~\ref{f1} and we have
    set
    $\Omega = \Gamma$.
    \label{evolfig}
 }
  \end{center}
\end{figure}

The time taken to reach this limit can be derived in the following
way \cite{mic06}.  The solution of Eq.~(\ref{vEoM}) is
\beq
  \mathbf{v}(t)= \mathbf{v}_\infty +
 e^{\mathbf{X} t}  \rb{\mathbf{v}_0-\mathbf{v}_\infty}
\eeq
with initial vector $\mathbf{v}_0 = \rb{1/2,1/2,0}$. In the long
time limit this can be approximated as
\beq
  \mathbf{v} \sim \mathbf{v}_\infty +
  \rb{\mathbf{v}_0-\mathbf{v}_\infty} e^{ -t/T_0}
  \label{vinfty}
\eeq
with the characteristic time defined through $T_0^{-1} =
\mathrm{min}\left\{ |\mathrm{Re}(\lambda_i)|\right\}$, where
$\left\{\lambda_i\right\}$ are the eigenvalues of matrix
$\mathbf{X}$, all of which have negative real parts.  This
characteristic time is found to be
\beq
  T_0 = \frac{3 \lambda^{1/3}}{\Gamma}
  \left[
    3^{2/3}(1-4r^2) + 3^{1/3} \lambda^{2/3} - 3 \lambda^{1/3}
  \right]^{-1}
\eeq
with
$
  \lambda = 9 r^2 +
  \sqrt{192r^6 - 63 r^4 + 36r^2 -3}
$
and $r = \Omega /\Gamma$.
In Fig.~\ref{T0fig}, we plot this characteristic time as a function
of the laser Rabi frequency in units of linewidth.
Figure~\ref{evolfig} shows a typical evolution of the system and
shows how well the behaviour of the full system is approximated by
Eq.~(\ref{vinfty}) with $T_0$ as above.

The characteristic time $T_0$ has the following simple limits
\beq
  T_0 =
  \left\{
    \begin{array}{cc}
      \Gamma / \Omega^2
        & \quad \mathrm{if}~ \Omega \ll \Gamma, \\
      2/ \Gamma
        & \quad \mathrm{if}~ \Omega \gg \Gamma. \\
   \end{array}
 \right.
\eeq
If the driving is weak $\Omega \ll \Gamma$ then the time to reach
the asymptotic population is slow. However, for laser amplitudes
greater than the relaxation rate, the characteristic time saturates
at a value twice that of the trion lifetime. This makes sense since,
in this limit, the speed of the system is limited by trion
relaxation, in which case, one half of the spin population is
transferred to $\ket{x-}$ in time $\Gamma^{-1}$, whence
$T_0=2\Gamma^{-1}$.  With the value $\Gamma=1.2~\mu$eV we obtain
$T_0 \approx 1.1$ns, which is far shorter that the hole-mixing
spin-flip transition time of $\gamma^{-1} \approx 1.6~\mu$s of
Ref.~\cite{ata06}.
Figure \ref{T0fig} also shows that this limit of $T_0 =
2\Gamma^{-1}$ is a good approximation for all $\Omega/\Gamma \gtrsim
1$.

We now consider the inclusion of the off-resonant transition, which
acts to reduce the asymptotic value of $\rho_{--}$ away from unity.
The full Hamiltonian of Eq.~(\ref{Ham1}) depends on time, and
therefore we cannot simply set $\dot{\rho}=0$ to find the asymptotic
solutions.
Rather, we proceed  by making the following Ansatz for the
asymptotic density matrix elements \cite{sto96}
\beq
  \rho_{ij}(t\rightarrow \infty) = \rho^{(0)}_{ij} + \sum_\pm \rho^{(\pm)}_{ij}
  e^{\pm i \Sigma_B t}
  ,
\eeq
where the coefficients $\rho^{(0,\pm)}_{ij}$ are stationary. We
place this Ansatz into Eq.~(\ref{master}) and neglect terms
oscillating as frequencies faster than $\Sigma_B$.  This results in
a set of algebraic equation for the coefficients
$\rho^{(0,\pm)}_{ij}$ which we simply solve. We obtain the following
expressions for the steady-state coefficients
\beq
  \rho_{x+,x+}^{(0)}&=& \frac{ \Gamma^2 + \Omega^2}{D}
  ;\nonumber\\
  \rho_{x-,x-}^{(0)} &=& 1 - \frac{ \Gamma^2 + 3\Omega^2}{D}
  \nonumber\\
  \rho_{\tau-,\tau-}^{(0)} &=& \rho_{\tau+,\tau+}^{(0)} =\Omega^2/D
  ;\nonumber\\
  \rho_{x+,\tau+}^{(0)} &=& i \Gamma \Omega/D
  ;\nonumber\\
  \rho_{x-,\tau-}^{(-)}  &=& \rb{\rho_{\tau-,x-}^{(+)}}^*
    = \Omega(i \Gamma -\Sigma_B)/D
  \label{rhoME}
    ,
\eeq
where the denominator $D=\Sigma_B^2 + 2 \Gamma^2+4\Omega^2$, and all
the other coefficients are zero.

Let us define the fidelity of the state preparation as
$F=\ew{\Psi|\rho|\Psi}$, where $\ket{\Psi}$ is the desired target
state with population localized in the state $\ket{x-}$ and $\rho$
is the actual density matrix of the final states.  This evaluates
simply as $F=\rho_{x-,x-}=\rho_{x-,x-}^{(0)}$ and starts with a
value of 1/2 in the initial unpolarized state, and is unity for
100\% purification  \cite{ata06a}.  Let us define as $\epsilon$ the
amount by which $F$ differs from unity: $\epsilon=1-F$.
From Eqs.~(\ref{rhoME}), we therefore find that the
state-preparation error is
\beq
  \epsilon
    = \frac{ \Gamma^2 + 3\Omega^2}{\Sigma_B^2 + 2\Gamma^2+4\Omega^2}
  \approx \frac{\Gamma^2 + 3 \Omega^2}{\Sigma_B^2}
  ,
\eeq
where we have made use of $|\Sigma_B| \gg \Omega,\Gamma$.

\begin{figure}[t]
  \begin{center}
  \psfrag{err}{$\log_{10}\epsilon$}
  \psfrag{r}{$\Omega/\Gamma$}
  \epsfig{file=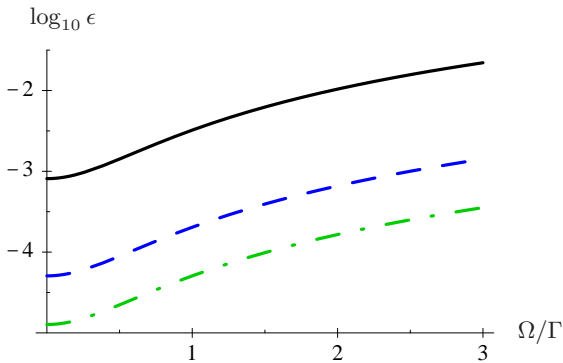, clip=true,width=0.9\linewidth}
  \caption{ The logarithm of the error $\epsilon=1-F$, with $F$ the
    fidelity,
    as a function of the laser Rabi energy $\Omega$.
    Results are shown for three different fields:
    1~T (black solid), 4~T (blue dash), and 8~T (green dash-dot).
    Other parameters as Fig~\ref{f1}.
    \label{errfig}
 }
  \end{center}
\end{figure}
%
In Fig.~\ref{errfig} we plot the full result for $\epsilon$ as a
function of the Rabi frequency. The most salient point is that for a
field of the order of 1T, and with $\Omega/\Gamma = 1$, the error
$\epsilon$ is of the order of $3\times10^{-3}$, which is very small,
and of the order of the measurement threshold described in
Ref.~\cite{ata06}. Increasing the field, decreases the error and at
a high laboratory field such as 8~T the error is reduced to
$\epsilon = 5\times10^{-5}$. These estimates agree very well with
the results of numerical integration of the equations of motion.
It should be noted that these values apply whilst the CW
illumination is still in effect. Turning off the laser allows
population trapped in the trion states to relax back to the
ground-state sector with rate $\Gamma$.  Half of this population
ends up in the required state $\ket{x-}$, reducing the error by a
factor of 2/3.

In summary then, we have considered the advantages of using the
Voigt configuration for the preparation of the spin-state of an
electron in a self-assembled QD.  Provided that the Rabi frequency
of the laser is greater than the trion relaxation rate, the state
preparation is fast, proceeding with a time-scale of $2\Gamma^{-1}
\approx 1$~ns, which is orders of magnitude faster than in the
Faraday configuration. Use of the Voigt configuration does, however,
impose an upper limit on the maximum obtainable fidelity, but this
is small, with the deviation from unity being $\epsilon \approx (
\Gamma^2 + 3 \Omega^2)/\Sigma_B^2 \approx 10^{-3}$ at 1T. This
approach therefore represents a fast way of initializing an electron
spin to high fidelities for quantum information processing.

This work was supported by ARO/LPS and by grant NSF DMR 0403465.  We
are grateful to M.~Atat\"{u}re for helpful discussions.



\begin{thebibliography}{99}
\bibitem{note_gs}  If the g-factors had opposite signs, we
    would have $|\Delta_B| > |\Sigma_B|$ and therefore driving one of the $H$
    transitions instead would be the best strategy to
    minimise initialisation error.
\bibitem{mix_note}  The dominant hole mixing here is between
    heavy hole $\ket{\frac{3}{2};\pm\frac{3}{2}}$
    and light hole $\ket{\frac{3}{2};\mp\frac{1}{2}}$.
    By adjusting the relative phase between circular-polarized components
    of the applied laser, the effect of this mixing
    can be reduced to only altering the Hamiltonian of Eq. (\ref{Ham1}) through a rescaling
    of the Rabi frequency.
\bibitem{ata06a}  In the limit that $1-\rho_{x-,x-} \ll 1$, this fidelity is
    approximately the same as the definition $F=1-\rho_{x+,x+}/\rho_{x-,x-}$
    used in Ref.~\cite{ata06} (M.~Atat\"ure, private
    communication).
\end{thebibliography}
\end{document}